\documentclass[aps,pra,preprint]{revtex4}
\usepackage{graphicx}% Include figure files
\usepackage{bm}
\usepackage{amsmath}
\begin{document}

\title{Microscopic nonequilibrium dynamics of an inhomogeneous Bose gas beyond the Born approximation}

\author{S. G. Bhongale, R. Walser, and M. J. Holland}

\affiliation{JILA and Department of Physics, University of Colorado,
  Boulder CO 80309-0440, USA.}

\date{\today}

%\Wideabs{

\begin{abstract}
  Using the prescription of the nonequilibrium statistical operator
  method, we derive a non-Markovian generalization to the kinetic
  theory described by Walser {\sl et al.} [Phys. Rev. A {\bf 59}, 3878
  (1999)]. Quasi-particle damping and effects arising from the finite
  duration of a collision are introduced to include terms beyond
  the Born approximation. Such a self-consistent theory is shown to
  conserve energy to second order in the interaction strength, even in
  the Markov limit. This kinetic theory is applied to a simple model
  of a Bose gas confined in a spherical trap to study the full
  real-time evolution towards equilibrium. A modified form for the
  damping function, is seen to strongly improve the energy
  conservation.  Based on a linear response calculation, we predict
  the damping rates and frequencies of the collective excitations. We
  demonstrate the emergence of differing time scales for damping and
  equilibration.
\end{abstract}

\maketitle
%}
\section{Introduction}
In the mean-field approximation, a Bose-condensed phase is well
described by the Gross-Pitaevskii (GP) equation \cite{dalfovo}.
Examples of collective effects that arise at the mean-field level
include the formation of vortex states
\cite{murray,haljan,raman,ketterle,dalibard} and collective
excitations \cite{jin96,edwards,stringari}. The GP equation is often
sufficient to describe the dynamics at $T=0$. However, at finite
temperatures, $T$ near $T_c$ [the Bose-Einstein condensation (BEC)
temperature], effects due to the presence of a thermal component may
be important. The hot thermal atoms interact with each other and with
the condensed atoms via binary collisions. In fact, collisional
dynamics is the essential mechanism for evaporative cooling leading to
BEC. Finite temperature effects generate phenomena such as phase
diffusion and damping of collective excitations. In order to describe
these effects, inclusion of collisions arising from the thermal
component becomes necessary. A generalized GP equation, for example
the Hartree-Fock-Bogoliubov approach, includes collisions only as
first-order energy shifts and is therefore valid only at low
temperature. In the other limit, $T>T_c$ (high temperature), the
condensate component may be negligible and the dynamics is then
dominated by the thermal component. This limit is well described by
the quantum-Boltzmann (QB) equation. A second order kinetic theory,
which reduces to the GP and the QB equations in the limits $T=0$ and
$T>T_c$, respectively, is necessary to provide a complete description.

Several different versions of quantum kinetic theories have been
formulated to describe a quantum degenerate Bose system
\cite{gardiner971,gardiner973,milena,walser,gardiner975,stoof,morgan,burnett}.
However certain questions remain unanswered. An important issue is the
question of Markovian versus non-Markovian dynamics in a
non-homogeneous system. Previous attempts to address this issue
\cite{morozov,morozov2,bonitz,kremp,bonitzbook}, have been limited to
the discussion of system with translational symmetry.

In the case of a sufficiently dilute nonhomogeneous system, one may
consider the neglection of memory effects under the principal of the
rapid attenuation of correlations. Such an ansatz will result in a
Born-Markov approximation. While in this limit kinetic equations are
vastly simplified, they do not conserve energy. Also, the Born-Markov
theory involves a regularizing factor $\eta\rightarrow 0$
\cite{zubarev,peletminskii}. Such a limiting procedure is
computationally nontrivial to implement numerically if one is dealing
with a finite system with discrete energy levels. If the resolution is
instead to include memory effects and initial correlations, then the
full non-Markovian theory which arises is described by complex kinetic
equations.

In this article we wish to address some of these issues by
generalizing the kinetic theory in Ref.~\cite{walser} to include
effects beyond the Born approximation such as quasi-particle damping
and effects arising from the finite duration of a collision. We
hypothesize the form of the damping function by looking at the
corresponding retarded Green's function \cite{haug,semkat}.  We show
that such a perturbation theory conserves energy up to second order in
the interaction strength.

We apply this theory to a simple model of a Bose gas confined in a
spherical box above $T_c$. The simplicity of this model allows us to
study the real-time (nonequilibrium) evolution of the system from
some initial state to a final state of equilibrium. The equilibrium
state is essentially a self-consistent steady-state solution of the
kinetic equation. We perform a linear response calculation in order to
verify the stability and to study the damping rates and frequencies of
the steady-state solution. We show the existence of different time
scales corresponding to damping and equilibration.

This article is organized as follows. In Sec.~\ref{kequation} we
derive a non-Markovian generalization to the kinetic theory of
Ref.~\cite{walser} using the prescription of a nonequilibrium
statistical operator method \cite{zubarev}. Both quasi-particle
damping and damping arising from the finite time of collision events
will be discussed in Sec.~\ref{approximations}. This has
implications on the underlying symmetries of the theory and their
associated conserved quantities. In Sec.~\ref{application}, we
apply this generalized kinetic theory to a simple model of a dilute
Bose gas confined in a spherical box and obtain a self consistent
steady-state solution to the second-order kinetic theory. Finally in
Sec.~\ref{realresponse}, we study the response of the system to a
small perturbation. This will help us to predict frequencies and
damping rates of collective modes.

\section{\label{kequation}Kinetic equations}
We start with the many-body Hamiltonian for a weakly interacting Bose
gas given by
\begin{equation}
\label{Htot}
\hat{H}=\hat{H}^{(0)}+\hat{H}^{(1)},
\end{equation} 
where $\hat{H}^{(0)}$ is the single particle Hamiltonian that is
defined as
\begin{equation}
\hat{H}^{(0)}=\varepsilon^{12} 
\,\hat{a}_1^{\dagger} \hat{a}_2^{\phantom \dagger}
\end{equation} 
using the implicit summation convention for repeated indices. The
two-body energy $\hat{H}^{(1)}$ is given by
\begin{equation}
\hat{H}^{(1)}=\phi^{1234}\, \hat{a}_1^{\dagger}\hat{a}_2^{\dagger}
\hat{a}_3^{\phantom \dagger} \hat{a}_4^{\phantom \dagger}.
\end{equation}
The bosonic operators $\hat{a}_1$ and $\hat{a}^{\dagger}_1$ annihilate
and create a particle in a single particle state $|1\rangle$,
respectively. The abbreviated notation $|1\rangle$ represents a state
specified by a complete set of quantum numbers for both the motional
and electronic degrees of freedom.

We assume that the particles are confined by an external trapping
potential $V_{\text{ext}}$.  Thus, the matrix elements of the single
particle Hamiltonian are given by
\begin{eqnarray}
  \varepsilon^{12}&=&\langle 1| \frac{\hat{{\bf p}}^2}{2m}
  +V_{\rm{ext}}(\hat{{\bf x}})|2\rangle,
\end{eqnarray}
where $m$ is the mass.  The binary interaction is mediated by a short
range repulsive potential $V_{\text{bin}}$. 
This gives
the symmetrized $({\cal S})$ matrix elements 
\begin{eqnarray}
  \phi^{1234}&=&({\cal S}) \langle 1|\otimes\langle 2| 
  V_{\text{bin}}(\hat{{\bf x}}\otimes {\bf \openone}-
    {\bf \openone}\otimes \hat{{\bf x}})
  |3\rangle \otimes |4\rangle.
\end{eqnarray}
In the low energy limit $V_{\rm{bin}}$ can be approximated by a contact potential with the matrix elements given by
\begin{eqnarray}
  \phi^{1234}&\approx&\frac{V_0}{2}\int_{-\infty}^{\infty} 
  \langle 1|{\bf x}\rangle\langle 2|
{\bf x}\rangle\langle {\bf x}|3\rangle\langle
  {\bf x}|4\rangle \,d^3{x}.
  \label{phis}
\end{eqnarray}
The interaction strength $V_0$, is related to the scattering length
$a_{\text{S}}$ by $V_0=4\pi\hbar^2a_{\text{S}}/m$.
  
Here, we use the well-known nonequilibrium statistical operator method
\cite{zubarev,peletminskii} to obtain an explicitly non-Markovian
version of the kinetic theory \cite{walser}. In this approach, the
nonequilibrium state of a weakly interacting quantum gas is specified
by a set of single time master variables. For our system the most
important master variable is the single particle density matrix
$f(t)$,
\begin{equation}
  f_{12}(t)=\langle \hat{a}_2^{\dagger}\hat{a}_1^{\phantom \dagger}\rangle=
  {\rm Tr}\left\{\hat{a}_2^{\dagger}\hat{a}_1^{\phantom \dagger} \,
    \sigma(t)\right\},  
\end{equation}
where $\langle\ldots\rangle=\text{Tr}\{\ldots\sigma(t)\}$ and
$\sigma(t)$ is the statistical many-body density operator. We focus
our studies on the temperature regime above and in close proximity to
the critical temperature for BEC. For this reason, we do not consider
either symmetry breaking fields, $\langle \hat{a}_1\rangle$, or the
anomalous fluctuations, $\langle \hat{a}_1\hat{a}_2\rangle$. We
therefore define $\{ \hat{\gamma}_{0}^{\phantom
  \dagger}=\openone,\hat{\gamma}_{k}^{\phantom
  \dagger}=\hat{a}_2^{\dagger}\hat{a}_1^{\phantom \dagger}|k\in
\{(1,2)\}\}$ as our complete set of relevant operators. The
expectation values of these operators, $\gamma_{k}(t)=\langle
\hat{\gamma}_{k} \rangle$, are the only quantities that will appear
in the final kinetic equations.
  
The time evolution of the nonequilibrium statistical operator
$\sigma(t)$ is described by the Liouville equation with an extra
source term on the right hand side:
\begin{equation}
  \frac{d}{d t}\sigma(t)+
  \frac{i}{\hbar}[\hat{H},\sigma(t)]=-\eta\,\left(\sigma(t)-\sigma^{(0)}(t)\right).
\label{liouville}
\end{equation}
Such a source term breaks the time reversal symmetry of the Liouville
equation and represents a convenient way to incorporate the
irreversible character of macroscopic processes. The relevant
distribution $\sigma^{(0)}(t)$ given by 
\begin{equation}
  \sigma^{(0)}(t)= 
  \sigma^{(0)}_{\{ \gamma(t) \}}= \exp{\left\{
     \hat{\gamma}_k \Upsilon^k(t)\right\}},
  \label{boundaryc}
\end{equation}
where $\Upsilon^k(t)$ are the Lagrange multipliers, represents a
special solution that maximizes the information entropy $S'=-{\rm
  Tr}\{\sigma'\log(\sigma')\}$ for the given averages $\gamma_k(t)$.
Furthermore, at some initial instance $t=t_0$ in the remote past, we
can assume that $\sigma(t_0)$ corresponds to its noninteracting value
and therefore
\begin{equation}
\sigma(t_0)=\sigma^{(0)}(t_0).
\end{equation}
The Lagrange multipliers $\Upsilon^k(t)$ are calculated from the
self-consistency condition
\begin{eqnarray}
  \gamma_k(t)&=&
  \text{Tr}\{ \hat{\gamma}_k \, \sigma^{(0)}(t) \}=
  \text{Tr}\{ \hat{\gamma}_k \, \sigma(t) \}.\label{sconsist}
  \label{averages}
\end{eqnarray}
This essentially enforces the Chapman-Enskog condition \cite{chapman} for the
restricted set of relevant operators at all times. 

From the Liouville
equation (\ref{liouville}) one
can easily establish the basic equations of motion for the
average values, 
\begin{eqnarray}
  \frac{d}{dt}\gamma_k(t)
&=&\frac{i}{\hbar}{\rm Tr}\Big\{ [\hat{H}^{(0)},\hat{\gamma}_k]\,\sigma(t)\Big\}\nonumber\\
&+&\frac{i}{\hbar}{\rm Tr}\Big\{ [\hat{H}^{(1)},\hat{\gamma}_k]\,\sigma(t)\Big\}.
  \label{relopereqn}
\end{eqnarray}
The form of $\hat{H}^{(0)}$ enables us to express the first trace on
the right hand side of the above equation in terms of the averages $\gamma_k(t)$.
The second trace plays the role of the ``collision'' term, evaluation of
which requires us to seek an integral solution of the Liouville
equation. But, before we
proceed, it is instructive to repartition the
total Hamiltonian Eq.~(\ref{Htot}) into single particle and two-particle
contributions,
\begin{eqnarray}
\hat{H}&=&\hat{\bar{H}}^{(0)}(t)+\hat{\bar{H}}^{(1)}(t)\nonumber\\
&=&\left[\hat{H}^{(0)}+\hat{Q}(t)\right]
+\left[\hat{H}^{(1)}
-\hat{Q}(t)\right].
\end{eqnarray}
 This modification anticipates
self-energy shifts
\begin{equation}
\hat{Q}(t)=Q^{12}(t)\hat{a}_1^{\dagger}\hat{a}_2^{\phantom \dagger}, 
\end{equation}
which will inevitably arise in the course of the calculation.

An integral solution for $\sigma(t)$ can then be obtained easily from
the Liouville equation (\ref{liouville}), by using the single particle
time evolution operator ${\protect{\hat{U}^{(0)}}}(t,t_1)$, the boundary
condition Eq.~(\ref{boundaryc}), and an additional partial integration.
Thus, one finds
\begin{eqnarray}
\sigma(t)&=&\sigma^{(0)}(t)-\int_{t_0}^t
dt_1\,e^{-\eta(t-t_1)}{\hat{U}^{(0)}}(t,t_1)\nonumber\\
&\times&\bigg[\frac{d}{d t_1} \sigma^{(0)}(t_1)
+\frac{i}{\hbar}[\hat{\bar{H}}^{(0)}(t_1),\sigma^{(0)}(t_1)]\nonumber\\
&+&\frac{i}{\hbar}[\hat{\bar{H}}^{(1)}(t_1),\sigma(t_1)]\bigg]
{\protect{\hat{U}^{(0)}}}{}^{\dagger}(t,t_1).\label{ieqn}
\end{eqnarray}
Since the Hamiltonian $\hat{\bar{H}}^{(0)}(t)$ depends on time through
the $\hat{Q}(t)$, the time evolution operator ${\hat{U}}^{(0)}$ is in
general a time-ordered exponent
\begin{equation}
{\hat{U}^{(0)}}(t,t_0) =T\exp\left[
-\frac{i}{\hbar}\int_{t_0}^t dt_1\,\hat{\bar{H}}^{(0)}(t_1) \right].
\label{propagator}
\end{equation}
To establish the time derivative $d\sigma^{(0)}(t_1)/dt_1$  in
Eq.~(\ref{ieqn}), we recall from Eq.~(\ref{boundaryc}) that the
relevant operator depends only implicitly on time
through the averages $\gamma_k(t_1)$. Moreover by exploiting the
transformation properties of a quantum Gaussian (see Appendix
\ref{appref}), one finds that
\begin{eqnarray}
  \frac{d}{d t}\sigma^{(0)}(t)&=&
    \frac{d}{dt}\gamma_k(t)\,\partial_{\gamma_k} \sigma^{(0)}(t)=
  -\frac{i}{\hbar}[\hat{\bar{H}}^{(0)}(t),\sigma^{(0)}(t)]\nonumber\\
  &+&
  \frac{i}{\hbar}{\rm Tr}\left\{ [\hat{\bar{H}}^{(1)}(t),\hat{\gamma}_k]
    \,\sigma(t)\right\}\,\partial_{\gamma_k} \sigma^{(0)}(t).
\label{eqx}
\end{eqnarray}
Using these relations we eliminate the time derivative of
$\sigma^{(0)}(t_1)$ in Eq.~(\ref{ieqn}), to obtain the integral form of the
statistical operator
\begin{eqnarray}
  \sigma(t)&=&\sigma^{(0)}(t)-\frac{i}{\hbar}\int_{t_0}^t dt_1
  e^{-\eta(t-t_1)}{\hat{U}^{(0)}}(t,t_1)\nonumber\\
  &\times&\bigg[\text{Tr}\left\{[
    \hat{\bar{H}}^{(1)}(t_1),{\hat \gamma}_k]\, \sigma(t_1)\right\}
  \partial_{\gamma_k}\sigma^{(0)}(t_1)\nonumber\\
  &+& [\hat{\bar{H}}^{(1)}(t_1),\sigma(t_1)]
  \bigg]{\hat{U}^{(0)}}{}^{\dagger}(t,t_1).
\end{eqnarray}
Since we are only interested in a weakly interacting gas, we seek a
power series expansion in the interaction strength
\begin{eqnarray}
\sigma(t)&=&\sigma^{(0)}(t) + \sigma^{(1)}(t)+...\\
\sigma^{(1)}(t)&=&-\frac{i}{\hbar}\int_{t_0}^t dt_1
  e^{-\eta(t-t_1)}{\hat{U}^{(0)}}(t,t_1)\nonumber\\
  &\times&\bigg[\text{Tr}\left\{[
    \hat{\bar{H}}^{(1)}(t_1),{\hat \gamma}_k]\, \sigma^{(0)}(t_1)\right\}
  \partial_{\gamma_k}\sigma^{(0)}(t_1)\nonumber\\
  &+& [\hat{\bar{H}}^{(1)}(t_1),\sigma^{(0)}(t_1)]
  \bigg]{\hat{U}^{(0)}}{}^{\dagger}(t,t_1).\label{sigmat}
\end{eqnarray}
With this explicit expression for the statistical operator, the evaluation of the equation of motion (\ref{relopereqn}) is straight forward and one obtains the quantum-Boltzmann equation
\begin{eqnarray}
\frac{d}{dt}f(t)&=&{\cal L}[f]+{\cal L}[f]^{\dagger}\label{fequation}.
\end{eqnarray}
Here, Wick's theorem has been used to express the higher order
averages in terms of the single particle ones.  The kinetic operator
${\cal L}$ consists of a reversible Hartree-Fock (HF) part ${\cal
  L}_{\rm{HF}}$ and a collisional quantum-Boltzmann contribution
${\cal L}_{\rm{QB}}$,
\begin{eqnarray} 
{\cal L}[f]&=&{\cal L}_{\text{HF}}[f]+{\cal L}_{\text{QB}}[f],\nonumber\\
{\cal L}_{\text{HF}}[f]&=&-\frac{i}{\hbar}H_{\text{HF}}(t) f(t),\nonumber\\
{\cal L}_{\text{QB}}[f]&=&\Gamma_{ff(1+f)(1+f)}-\Gamma_{(1+f)(1+f)ff}.\nonumber
\end{eqnarray}
$H_{\rm{HF}}=\varepsilon+2\,U_f$ is the Hartree-Fock Hamiltonian, with
$U_{f}^{14}=2\,\phi^{1234}f_{32}$ the self energy, while $\Gamma$'s
are the collision integrals given by
\begin{eqnarray}
  \Gamma_{ABCD}^{15}&=&\frac{1}{\hbar}\int_{t_0}^t dt_1\bigg[
  e^{-\eta(t-t_1)}\phi^{1 2 3 4}
  \phi^{1^{\prime \prime}2^{\prime \prime}
    3^{\prime \prime} 4^{\prime \prime}}\nonumber\\
  &\times&
  {\cal K}_{1^{\prime \prime} 1^\prime}(t,t_1)
  {\cal K}_{2^{\prime \prime} 2^\prime}(t,t_1)
  {\cal K}_{3^{\prime \prime} 3^\prime}^{\dagger}(t,t_1)
  {\cal K}_{4^{\prime \prime} 4^\prime}^{\dagger}(t,t_1)\nonumber\\
  &\times&
  A_{31^\prime}(t_1)
  B_{42^\prime}(t_1)
  C_{4^\prime 2}(t_1)
  D_{3^\prime 5}(t_1)\bigg].
  \label{nonmarkovform}
\end{eqnarray}
The propagators ${\cal K}(t,t_0)$ are given by
\begin{equation}
\label{spprop}
{\cal K}(t,t_0)=
T\exp\left[-\frac{i}{\hbar}\int_{t_0}^t dt_1 \,H_{\rm{HF}}(t_1) \right].
\end{equation}
Therefore, unlike the collision terms of Ref.~\cite{walser}, the $\Gamma$'s
defined above depend on the past history of the system. Thus
Eq.~(\ref{fequation}) represents a non-Markovian generalization of the
kinetic equation previously derived in Ref.~\cite{walser}.

\section{\label{approximations}Conservation laws and Quasi-particle damping}
The conserved quantities for a closed isolated system are the total
energy $E$ and the total number $N$. These quantities then represent
the constants of motion for the full kinetic equation. The total
number operator can be represented as a linear combination of the
relevant operators,
\begin{equation}
\hat{N}=\hat{a}^{\dagger}_1\hat{a}_1^{\phantom \dagger},
\end{equation}
and therefore the functional $N(f)$, representing the total number is given by
\begin{equation}
N(f)={\rm Tr}\{f\}.
\end{equation}
The kinetic equation for $N(f)$ can then be written as
 \begin{eqnarray}
\frac{\partial}{\partial t}N(f)&=& {\rm Tr}\left\{{\cal L}[f]+{\cal L}[f]^{\dagger} \right\}\nonumber\\
&=&{\rm Tr}\left\{{\cal L}_{HF}[f]+{\cal L}_{HF}[f]^{\dagger} \right\}\nonumber\\
&+&{\rm Tr}\left\{{\cal L}_{QB}[f]+{\cal L}_{QB}[f]^{\dagger} \right\}.\label{Nequation}
\end{eqnarray}
The first order term on the right-hand side involves the trace of a
commutator and is trivially equal to zero. The $\Gamma$'s associated
with the second order terms have the following property due to the
symmetries of $\phi$,
\begin{equation}
{\rm Tr}\left\{\Gamma_{ff(1+f)(1+f)}\right\}={\rm
Tr}\left\{\Gamma_{(1+f)(1+f)ff}\right\}^*. 
\end{equation}
As a result, the second order contribution in Eq.~(\ref{Nequation}) can
also be shown to be zero. Hence the total number $N$ is a constant of
motion. The important point here is that the total number conservation is a
result of the symmetries of $\phi$ and does not dependent on the
non-Markovian nature of the collision integral. Therefore a Markov
approximation would leave this conservation law unchanged.

While the number conservation is a natural consequence of the
self-consistency condition, the total energy conservation is not
obvious as the Hamiltonian $\hat{H}$ cannot be represented as a linear
combination of the relevant operators. We start with writing the total
energy functional $E(f)$ as a perturbative expansion in
$\phi$,
\begin{eqnarray}
E(f)&=&{\rm Tr}\{\hat{H}^{(0)}\sigma^{(0)}\}\nonumber\\
&+&\Big[{\rm Tr}\{\hat{H}^{(1)}\sigma^{(0)}\} +{\rm Tr}\{\hat{H}^{(0)}\sigma^{(1)}\}\Big]\nonumber\\
&+&\Big[{\rm Tr}\{\hat{H}^{(1)}\sigma^{(1)}\} +{\rm Tr}\{\hat{H}^{(0)}\sigma^{(2)}\}\Big]+ ...
\end{eqnarray}
Using the self-consistency condition Eq.~(\ref{sconsist}) for the relevant operators,
one can show that the third and the fifth trace terms in the right-hand side of
the above expression drop out. The kinetic equation for the energy
functional $E(f)$ can then be written as
\begin{eqnarray}
\frac{\partial}{\partial t}E(f) &=& {\rm Tr}\left\{\varepsilon\dot{f}+U_f\dot{f}+U_{\dot{f}}f \right\}\\
&+&\frac{i}{2}{\rm Tr}\left\{\frac{\partial}{\partial t}\left[\Gamma_{ff(1+f)(1+f)}-\Gamma_{(1+f)(1+f)ff} \right]\right\}.\nonumber\label{kepluspe} 
\end{eqnarray}
Again, we use the symmetry properties of $\phi$, to write the simplified equation
\begin{eqnarray}
\frac{\partial}{\partial t}E(f)&=&-i\frac{\eta}{2}{\rm Tr}\left\{\Gamma_{ff(1+f)(1+f)}-\Gamma_{(1+f)(1+f)ff} \right\}.\nonumber\\\label{colle}
\end{eqnarray}
One can now see that the energy is conserved only in the $\eta
\rightarrow 0$ limit. In this limit the kinetic equation
(\ref{fequation}) represents the Born approximation. A finite value of
$\eta$ could then be thought of as resulting in additional terms that
are beyond the Born approximation. Such terms model the duration of
collision effects and quasi-particle damping. In principle if this
effect is treated self-consistently, $\eta$ will be a time dependent
function at least of order $\phi$. This means that the rate of change of
$E$ given by Eq.~(\ref{colle}) is of the order $\phi^3$.

Also, if one is only interested in times grater than the correlation
time $\tau_{{\rm cor}}$, the finiteness of $\eta$ allows us to extend
the lower limit of the collision integral to $-\infty$. Now we can
approximate the $f(t')$ in the non-Markovian expression of $\Gamma$ by
its instantaneous value $f(t)$ to obtain the Markov form
\begin{eqnarray}
\Gamma_{ABCD}^{(m)^{15}}&=&\frac{1}{\hbar}\int_{-\infty}^tdt_1\bigg[e^{-\eta(t-t_1)}\phi^{1234}\phi^{1''2''3''4''}\nonumber\\
&\times&{\cal K}_{1''1'}(t,t_1){\cal K}_{2''2'}(t,t_1){\cal K}_{3''3'}^{\dagger}(t,t_1){\cal K}_{4''4'}^{\dagger}(t,t_1)\nonumber\\
&\times&A_{31'}(t)B_{42'}(t)C_{4'2}(t)D_{3'5}(t)\bigg]\label{markovform}
\end{eqnarray}

One can verify that the above Markov form results in an energy
conservation, up to the most significant order given by
\begin{eqnarray}
\frac{\partial}{\partial t}E(f)&=&{\rm Tr}\left\{H_{\rm{HF}}\left({\cal L}_{\rm{QB}}^{(m)}[f]+{\cal L}_{\rm{QB}}^{(m)}[f]^{\dagger}\right)\right\}.\label{econsmarkov}
\end{eqnarray}
If we compare this with the expression for the correlation energy
$E_{\rm{cor}}$ in Refs.~\cite{morozov,morozov2}, the right hand side of
Eq.~(\ref{econsmarkov}) is exactly $-\partial E_{\rm{cor}}/\partial t$.
This is not surprising and can be understood more intuitively by
writing the collision integral as a sum of two contributions:
correlation and collision
\begin{equation}
\Gamma=\Gamma_{\rm{cor}}+\Gamma_{\rm{col}}=\int_{-\infty}^0...dt'+\int_0^t...dt'.
\end{equation}
For a finite $\eta$, the $\Gamma_{\rm{cor}}$ contribution decays
to zero as $e^{-\eta t}$. Therefore the decaying correlation energy
$E_{\rm{cor}}$ associated with this part shows up in the rate of
change of the total energy $E$. In the Born approximation
$\Gamma_{\rm{cor}}$ is constant because $\eta \rightarrow 0$.

The exponential damping results in a widened delta function and
therefore the rate of change of $E(f)$ Eq.~(\ref{econsmarkov}) can be
shown to be of order $\eta \Gamma^{(m)}$.  Thus, by including terms
beyond the Born approximation, we have obtained a collision integral
that is Markovian and still conserves energy up to $\phi^2$ order. Now
if we assume that the equilibration time is of the order
$1/\Gamma^{(m)}$ , the total change in energy $\Delta E$ is
therefore
\begin{eqnarray}
\Delta E&=&E(f^{\rm{eq}})-E(f^{\rm{in}}) \nonumber\\
&\sim& \eta\,\Gamma^{(m)}\times\left(\frac{1}{\Gamma^{(m)}}\right)\,=\, \eta, 
\end{eqnarray}
where $f^{\rm{in}}$ and $f^{\rm{eq}}$ are the initial and equilibrium
distributions, respectively.

The importance of the damping term becomes more obvious when one
attempts to solve the kinetic equation numerically. One no longer has
to worry about the $\eta\rightarrow 0$ limit in the Born-Markov
approximation, which for a finite system with discrete levels will
result only in exchange collisions and hence no equilibration.

\section{\label{application} 
  Application to a dilute Bose gas in a spherical box trap} In the
previous section, we introduced the general methods and concepts to
describe a weakly interacting Bose gas under nonequilibrium
conditions.  We will now apply these to a simple model of a typical
$^{87}$Rb experiment, as realized by many laboratories around the
world, for example Refs.~\cite{cornell,jin96,jin97}. The physical
parameters are usually quoted in the natural units for a harmonic
oscillator trapping potential, i.\thinspace{}e., the angular frequency
$\omega=2\pi\,200\,\text{Hz}$, the atomic mass $m_{87}=86.9092 \,
\text{amu}$, the ground state size $a_{\text{H}}=[\hbar/(\omega \,
m_{87})]^{1/2}= 763\,\text{nm}$, and the s-wave scattering length
$a_{\text{S}}=5.82\,\text{nm}$.

However, in the present article, we do not pursue the usual harmonic
confinement, but rather explore the properties of a radial box as a
particle trap. This choice is motivated by previous studies of the
self-consistent Hartree-Fock single-particle states \cite{walser2}.
As soon as repulsive mean-field potentials are added to the bare
harmonic trapping potentials, the corresponding eigenstates widen in
size and look remarkably close to the eigenstates of a box, provided
the spatial extensions of the box is chosen appropriately.  In particular,
we pick a box of radius $R=1000\, a_{\text{S}}=5.82\,\mu{}\text{m}$.

Thus, our model is represented by a spherical trap with box potential
given by
\begin{equation}
V_{\rm{ext}}(r)=\left\{
\begin{array}{cl}
  0,&r < R\\
  \infty,&r\ge R,
\end{array}
\right.
\end{equation}
The eigenfunctions are a product
of spherical Bessel functions $j_{(l)}$ and spherical
harmonics $Y_{(lm)}$:
\begin{figure}[t]
\includegraphics[scale=.35,angle=-90]{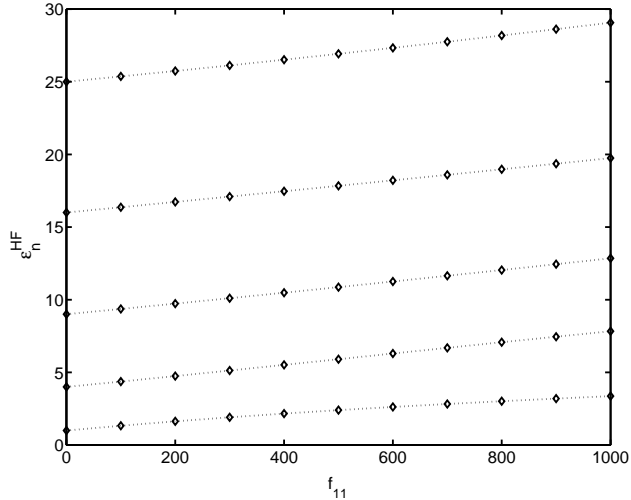}
\caption{Hartree-Fock energies as a function of particles number in
the box ground state, $f_{11}$ with all other $f_{ij}$'s equal to
zero.}
\label{energyshift}
\end{figure}
\begin{equation} 
\psi_{(nlm)}(r,\theta,\phi)=\left\{
\begin{array}{ll}
{\cal N}_{(nl)}\,j_{(l)}(r\, k_{(nl)})\,Y_{(lm)}(\theta,\phi),& r< R\\ 
0,& r\ge R.
\end{array}\right.
\end{equation}
A normalization constant is represented by ${\cal N}_{(nl)}$. The
eigenenergies are given in terms of the wave-vectors $k_{(nl)}$, which
can be obtained from the $n$-th nodes of the spherical
Bessel-functions of angular momentum $l$,
\begin{eqnarray}
  \varepsilon_{(nl)}&=& ( R\, k_{(nl)}/\pi)^2\,\varepsilon_0.
\end{eqnarray}
Here,  $\varepsilon_0$ represents the ground state energy and defines
the energy scale of the problem
\begin{eqnarray}
  \varepsilon_0&=&\frac{\hbar^2 \pi^2}{2\,m R^2}.
\end{eqnarray}
All the physical parameters are scaled with respect to this
energy unit $\varepsilon_0$ and the radius of the box $R$. 

For simplicity, we assume all the atoms to be in the $l=0$ state
initially. One therefore needs to consider only the $l=0$ manifold and
get for the normalization constant ${\cal N}_{(n0)}=n (\pi/(2
R^3))^{1/2}$ and radial wave-vector $R \,k_{(n0)}=n\pi$.  If the cloud
is relatively cold, then most of the population resides in the lowest
few energy states.  Therefore, we can also limit the number of radial
modes $1\leq n \leq n_{\text{max}}$.  For the present case, we take
$n_{\text{max}}=5$. Obviously, all these simplification reduce the
number of degrees of freedom significantly and thus we are able to
study certain aspects of the nonequilibrium dynamics of the trapped
Bose gas in great detail.

With the above definitions, the bare single particle box Hamiltonian
is given by
\begin{equation} 
\varepsilon^{pq}=\delta_{pq}\,q^2 \,\, (\text{no implicit sum over}\, q).
\end{equation}
The interaction part $\hat{H}^{(1)}$ involves the matrix elements of
the interaction potential defined by Eq.~(\ref{phis})
\begin{eqnarray}
\label{maxel}
\phi^{pqrs}&=&\frac{4 a_{\text{S}}}{\pi}
\int_0^\pi\sin{(p\, x)}\sin{(q\, x)}\sin{(r \,x)}\sin{(s\, x)}  
\frac{dx}{x^2},\nonumber\\
\end{eqnarray}
which in general have to be computed numerically. Interestingly,
in the case of a spherical box this integral can be evaluated
analytically and simplified to a finite sum of sine integrals and
cosine functions (see Appendix \ref{appmaxel} for details).

The nonequilibrium state of the above system is represented by the
single particle distribution function $f$ with time dependence given
by Eq.~(\ref{fequation}). Neglecting the second order collision terms,
the first order evolution is governed by the Hartree-Fock Hamiltonian
$H_{\rm{HF}}$ and is given by
\begin{equation}
\frac{\partial}{\partial t}f={\cal L}_{HF}[f]+{\cal L}_{HF}[f]^{\dagger}\label{firstorder}.
\end{equation}
The energy eigenstates of the interacting system are therefore
shifted from the bare box states due to the self energy effect. These
shifts can be significant depending on the total particle number. This
is clear from Fig.~\ref{energyshift} where we plot the eigenenergies
as a function of $f_{11}$, the total particle number in the box
ground state (with all other $f_{ij}$'s equal to zero).

\begin{figure}[t]
\includegraphics[scale=.35,angle=-90]{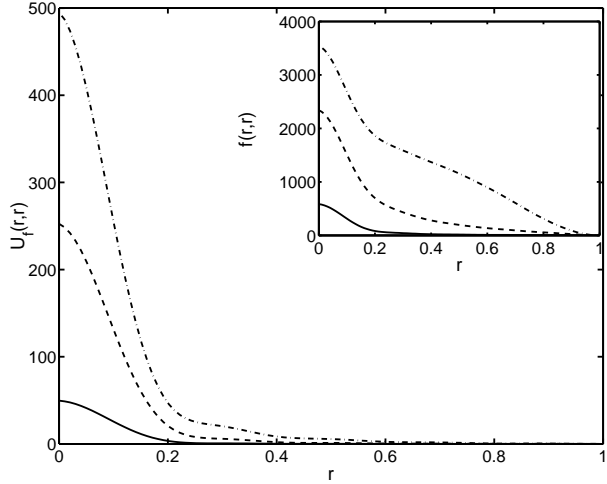}
\caption{The mean-field density, $U_f(r,r)$, as a
  function of the radial distance at a relatively hot temperature,
  $\beta=0.01$ is shown for three different values of particle
  numbers: $N=10$ (solid curve), $N=100$ (dashed), and $N=500$
  (dot-dashed). Inset shows the corresponding number density as a
  function of radial distance.}
\label{Ufsposinsethot}
\end{figure}

\begin{figure}[t]
  \includegraphics[scale=.35,angle=-90]{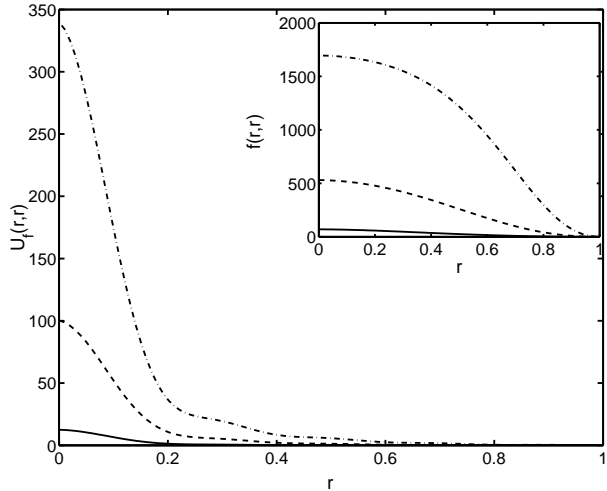}
\caption{The mean-field density, $U_f(r,r)$, as a
  function of the radial distance at a relatively cold temperature,
  $\beta=0.5$ is shown for three different values of particle numbers
  $N=10$ (solid curve), $N=100$ (dashed), and $N=500$ (dot-dashed).
  Inset shows the corresponding number density as a function of radial
  distance.}
\label{Ufsposinsetcold}
\end{figure}

Now note that the time dependent Hartree-Fock equation (\ref{firstorder}) for the density matrix, $f$, is nonlinear and hence we seek a self consistent
solution such that
\begin{equation}
f=\sum_i P(\varepsilon_i)|\varepsilon_i\rangle\langle\varepsilon_i|,
\end{equation}
where
$H_{\rm{HF}}|\varepsilon_i\rangle=\varepsilon_i|\varepsilon_i\rangle$,
and for a Bose-Einstein distribution $P(\varepsilon)$ is given by
\begin{equation}
P(\varepsilon)=\frac{1}{\exp((\varepsilon-\mu)/k_BT)-1}.
\end{equation}
For a given total particle number, $N$, and temperature, $\beta
=1/k_BT$, a self-consistent chemical potential, $\mu(\beta,N)$, and
hence a self-consistent Bose-Einstein distribution is obtained. For
example, let us consider three different total particle numbers
$N=10,100,500$ at two different temperatures, $\beta=1/k_BT=0.01,0.5$,
corresponding to hot and cold clouds, respectively.  In
Figs.~\ref{Ufsposinsethot} and \ref{Ufsposinsetcold}, we plot the
self-consistent solution and the self-energy density in the position
space representation. We see that the self-energy density is
proportional to the number density only near the center of the trap
and drops off faster with increasing radius. This can be attributed to
the restricted number of basis states used in our calculation and
effectively gives a finite range to the two body potential
$V_{\rm{bin}}$.

Up to first order, the $f$ equation is totally reversible. The
inclusion of the second order terms (collisions) break the
reversibility, and therefore represents a relaxation of the system
from some initial state to a final equilibrium state. Here we will be
using the Markov form (\ref{markovform}) for the collision integral.
We take the Hartree-Fock self-consistent state for the initial
condition.

\begin{figure}[b]
\centering\includegraphics[scale=.35,angle=-90]{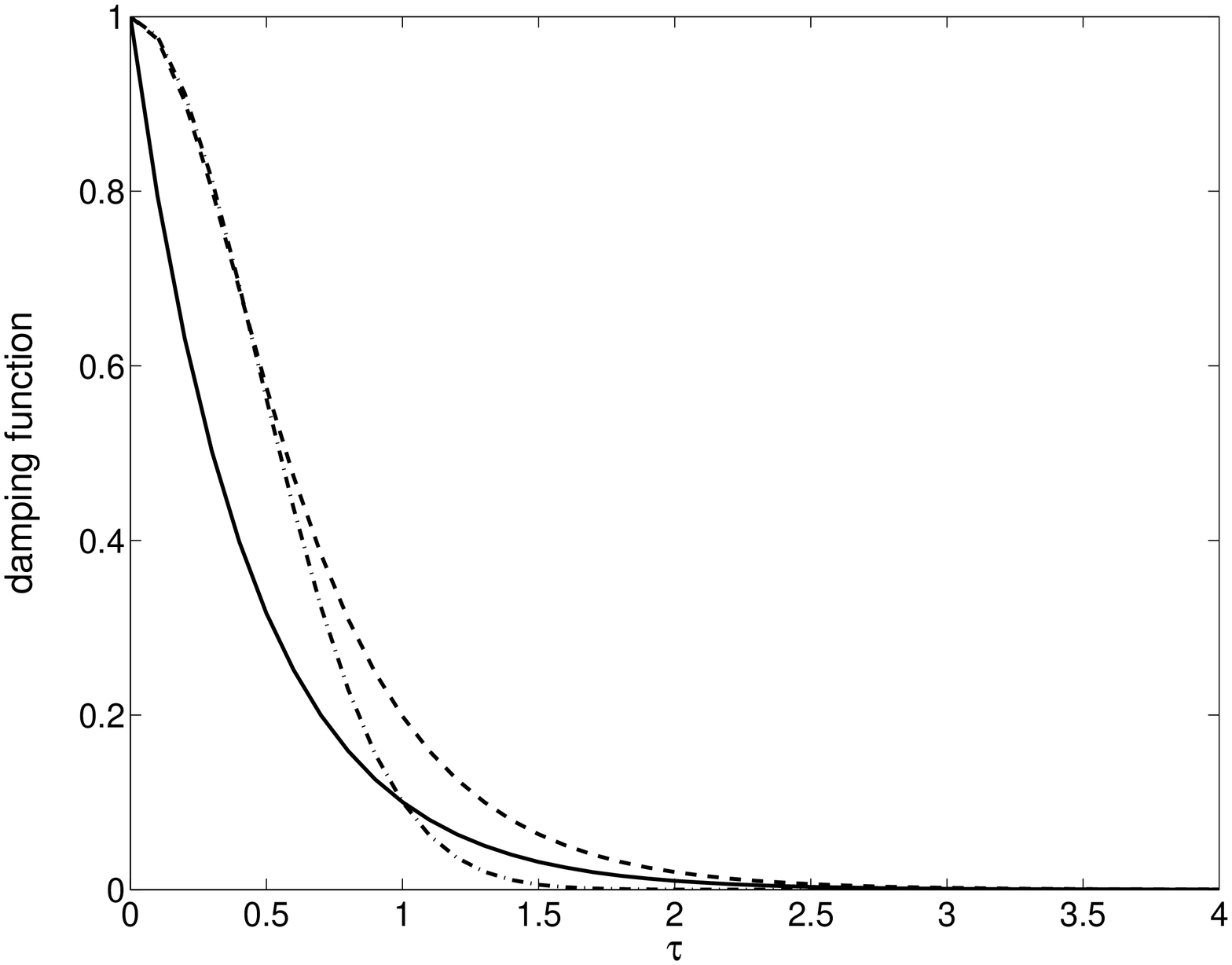}
\caption{\label{dampingfn}Comparison between different damping
  functions, $\exp(-\eta\tau)$ (solid), $\exp(-\eta\tau^2)$
  (dot-dashed), and $1/\cosh(\eta\tau)$ (dashed). Note that the
  hyperbolic secant function asymptotes to an exponential form for
  large $\tau$ and a Gaussian form for small $\tau$.}
\end{figure} 

In the previous section, we interpreted the function $\exp(-\eta\tau)$
to account for duration of collision effects and quasi-particle
damping. But the exponential form was originally introduced to break
the time reversal symmetry and it has the correct long time behavior.
However, an exponential damping will result in a Lorenzian line shape
for the final equilibrium distribution. Due to the long-reaching wings
of the Lorenzian curve, in the Markov limit, off-the-energy shell
collisions get weighted strongly. To seek an improved damping function
that will have the correct short and long time behavior, we use the
equivalence of kinetic theories based on the Green's function approach
\cite{kadanoff} and the non-equilibrium statistical operator method
\cite{zubarev} as shown in Ref.~\cite{wachter}.  The behavior of the
retarded Green's function $g(t,t_1)$ for very large and very small
time scales is given by
\begin{equation}
g(t,t_1)\sim\left\{\begin{array}{ll}e^{-\eta(t-t_1)}&t-t_1\gg\tau_{\rm{cor}}\\
e^{-\eta(t-t_1)^2}&t-t_1\ll\tau_{\rm{cor}}. \end{array}\right.
\end{equation}
Therefore, behavior over the intermediate time scale will be best
represented by an interpolating function. This is also true for the
damping function. From Fig.~\ref{dampingfn} we see that
the function 
\begin{equation}
{\cal F}(t,t_1)=1/\cosh(\eta(t-t_1)),
\end{equation}
has exactly this behavior and therefore represents a better choice
than the exponential form.

\begin{figure}[t]
\includegraphics[scale=.35,angle=-90]{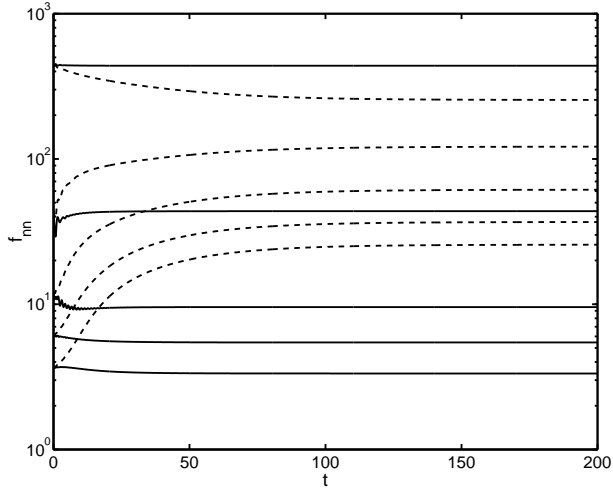}
\caption{Evolution of $f$ toward a self-consistent steady-state
  solution starting from an approximate solution. Two different
  damping functions are used in the evaluation of the $ \Gamma$'s.
  Dashed and solid curves correspond to exponential and hyperbolic secant
  respectively.}
\label{N500dynexp}
\end{figure}

\begin{figure}[t]
\centering\includegraphics[scale=.35,angle=-90]{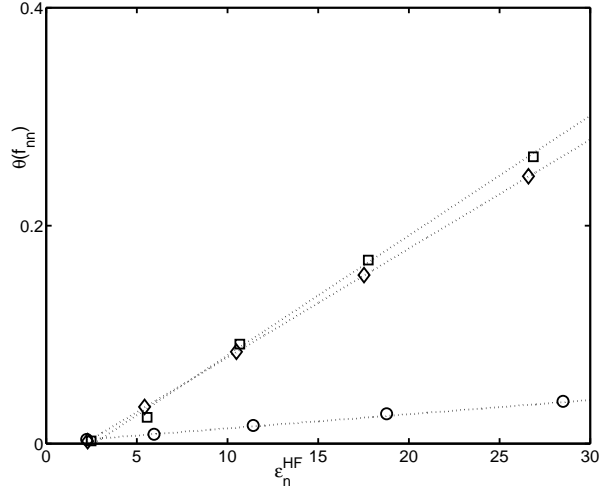}
\caption{\label{linearfit}Linear behavior of
  $\theta(f_{nn})$ as a function of $\varepsilon^{HF}_n$. Initial
  distribution shown with diamonds. Final distribution shown with
  circles for the case of exponential damping function, and squares
  for the case of hyperbolic secant damping function.}
\end{figure}

With either choice of the damping function, for a particular value of
the parameter $\eta$, a time propagation results in a self-consistent
steady-state solution, $f^{{\rm eq}}$. Figure \ref{N500dynexp} shows
such a time evolution for $N=500$ particles with an initial
temperature corresponding to $\beta=0.01$. We have chose $\eta$ to be
of the order ${\rm Re}[\Gamma]\approx 2.3$. As mentioned earlier we
see that the exponential damping function results in significant
transfer of population to the excited states. This effect is less with
the hyperbolic secant damping function. We get different steady-state
solutions because different damping functions correspond to different
initial correlations. Also a plot of
$\theta(f_{nn})\equiv\ln(1/f_{nn}+1)$ vs the Hartree-Fock energies
$\varepsilon^{HF}_n$, shown in Fig.~\ref{linearfit} shows that the
$f^{{\rm eq}}$ is very close to a Bose-Einstein distribution. The
slope, which represents the self-consistent value of $\beta$, shows
that the change in temperature is far greater for the exponential
damping function compared to the hyperbolic secant case, because
off-the-energy shell effects are larger as explained previously.

\begin{figure}[b]
\centering\includegraphics[scale=.35,angle=-90]{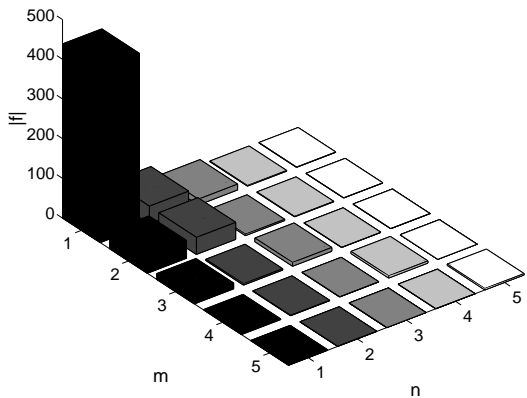}
\caption{Absolute value of $f^{{\rm eq}}$, the self consistent steady-state
  solution for the second order kinetic equation. The single particle
  density matrix is plotted in the Hartree-Fock basis.}
\label{fnewbasis}
\end{figure}

Thus we have obtained a self-consistent steady-state solution to the
second order kinetic equation. We emphasize here that the steady-state
solution is a result of the real-time non-equilibrium evolution of the
system. Deriving such an equilibrium solution, whose absolute value is
plotted in Fig.~\ref{fnewbasis}, is a prerequisite step to the study
of collective modes and damping rates of a dilute gas.

\section{\label{realresponse}Real time response to perturbation}
The properties of the equilibrium solution $f^{\text{eq}}$ exhibit the
expected characteristics of a Bose-Einstein distribution. In order to
verify the stability of this solution and to study the damping rates
of the collective excitations, we will now examine the real-time
response of the system to a perturbation.  First, we will outline the
linear response theory and discuss the structure of the modes, their
frequencies, and the life-time of the excitations.  Subsequently, we
will use these modes to initially prepare the system and to evolve the
full nonlinear quantum kinetic equation towards equilibrium.

One of the fundamental properties of the quantum kinetic equation
(\ref{fequation})
\begin{eqnarray}
\frac{d}{dt}f(t)&=&{\cal L}[f]+{\cal L}[f]^\dagger,
\end{eqnarray}
is its Hermitian structure. Thus, if we prepare a physical state
initially, it will remain Hermitian with $f(t)=f(t)^\dagger$,
indefinitely.  We will now consider a weak perturbation of an
equilibrium state,
\begin{eqnarray}
f(t)&=&f^{\text{eq}}+\delta f(t),
\end{eqnarray} 
and calculate the first order response of the system.
As usual, we want to assume that we can decompose a general
perturbation into fundamental damped and/or oscillatory eigenmodes of
the system.  Therefore, such a specific perturbation can be
parameterized as

\begin{figure}[t]
\centering\includegraphics[scale=.35,angle=-90]{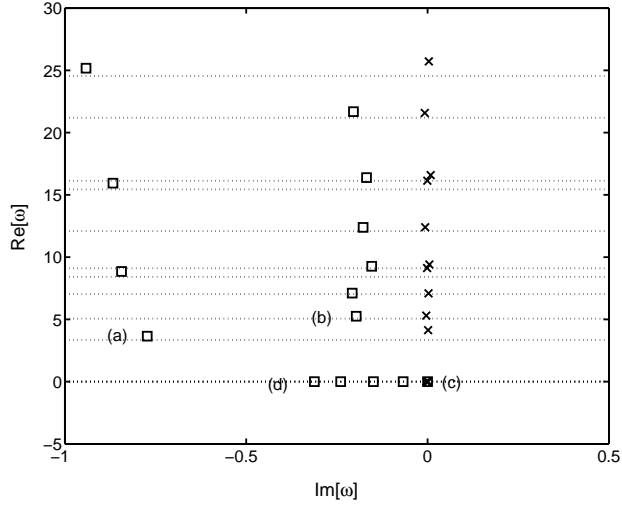}
\caption{Non-negative frequency eigenvalues shown with crosses for the Hartree-Fock equation and squares for the quantum-Boltzmann equation. The
  dotted lines correspond to difference energies
  $\varepsilon^{HF}_i-\varepsilon^{HF}_j$. The modes labelled (a) and
  (b) -- non-zero frequency and damping rate, (c) -- zero mode, and
  (d) -- zero frequency and non-zero damping rate will be considered
  for further discussion.}
\label{responseeigs}
\end{figure} 
\begin{eqnarray}
\delta f(t)&=&e^{-i\,\omega t}\delta f^{(+)}_\omega+\text{h.c.},\\
\delta f^{(+)}_\omega&=&\delta f^{(c)}_\omega+i\,\delta f^{(s)}_\omega,
\end{eqnarray}
where $\delta f^{(+)}_\omega$ denotes a positive frequency amplitude.  It
turns out to be useful to decompose it into quadrature
components
\begin{eqnarray}
\delta f^{(c)}_\omega&=&\frac{1}{2}\left[
{\delta f^{(+)}_\omega}^\dagger+\delta f^{(+)}_\omega\right],\\
\delta f^{(s)}_\omega&=&\frac{1}{2\,i}\left[
{\delta f^{(+)}_\omega}^\dagger-\delta f^{(+)}_\omega\right],
\end{eqnarray}
and evaluate the kinetic operator ${\cal L}[f]$ only for such
Hermitian arguments.  From a Taylor series expansion of the kinetic operator
around the equilibrium distribution, one obtains finally the linear
response equations for the fundamental modes
\begin{eqnarray}
(-i \omega)\,\delta f^{(+)}_\omega&=&
{\cal L}^{(1)}[\delta f^{(c)}_\omega]
+i\,{\cal L}^{(1)}[\delta f^{(s)}_\omega],\nonumber\\
&+&{\cal L}^{(1)}[\delta f^{(c)}_\omega]^\dagger
+i\,{\cal L}^{(1)}[\delta f^{(s)}_\omega]^\dagger,\label{eigeneqn}
\end{eqnarray}
where we have defined a linear response operator through an
appropriate centered difference limit

\begin{eqnarray}
{\cal L}^{(1)}[\delta f]&=&\lim_{\lambda\rightarrow 0}
\frac{ 
 {\cal L}[f^{\text{eq}}+\lambda \delta f]
-{\cal L}[f^{\text{eq}}-\lambda \delta f]}{2\lambda}.
\end{eqnarray}

\begin{figure}[t]
\centering\includegraphics[scale=.35,angle=-90]{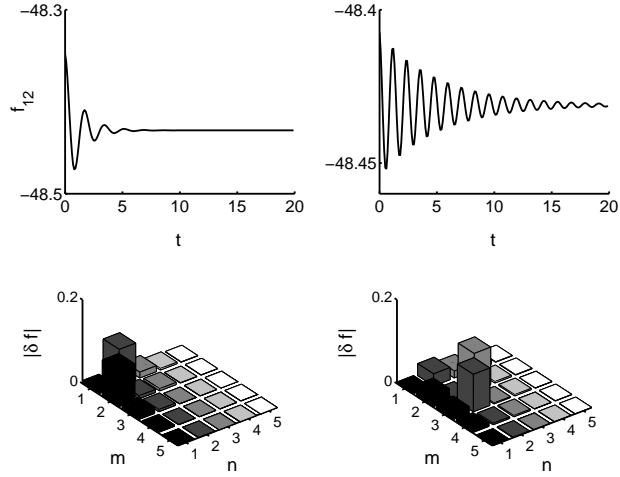}
\caption{The perturbation $\lambda \delta f^s_{\omega}$ (bottom)
  is shown in a rotated frame such that $f^{{\rm eq}}$ is diagonal; the
  resulting oscillatory and damped behavior (top) of the element $f_{12}$ of
  $f$ in the box basis is due to the perturbation. The left and right
  figures correspond to the points marked (a) and (b) in
  Fig.~\ref{responseeigs} respectively.}
\label{linresponse1}
\end{figure}

We solve Eq.~(\ref{eigeneqn}) as an eigenvalue problem. In general,
the eigenvalues are complex, with frequency and damping rate given by
the real and the imaginary parts respectively. The eigenvalues appear
as complex conjugate pairs. The eigenmodes corresponding to non-zero
eigenvalues are Hermitian conjugates of each other and are traceless
with normalization given by
\begin{equation}
{\rm Tr}\{ \delta f^{(+)}_{\omega}\delta f^{(+)}_{\omega}{}^{\dagger}\}=1.
\end{equation}
The physical linear response mode is given by the quadrature
components $\delta f^s_{\omega}$ and $\delta f^c_{\omega}$. There also
exists a zero mode that has nonvanishing trace. The damping rates
are all negative, thus confirming the stability of the collective
modes.  In Fig.~\ref{responseeigs}, we plot the positive frequency
eigenvalues.  The dotted lines correspond to the difference
frequencies of the Hamiltonian $H_{HF}$.

It is interesting to see how these different modes evolve in real
time.  For this we use the equilibrium distribution obtained in the
previous section and perturb it with one of the quadrature
components 
\begin{eqnarray}
f\rightarrow f^{{\rm eq}}+\lambda \delta f^{(s)}_{\omega},
\end{eqnarray}
where $\lambda=0.2$ determines the smallness of the perturbation.  In
particular we will consider the modes labeled by (a), (b), (c) and (d)
in Fig.~\ref{responseeigs}.

\begin{figure}[t]
\centering\includegraphics[scale=.35,angle=-90]{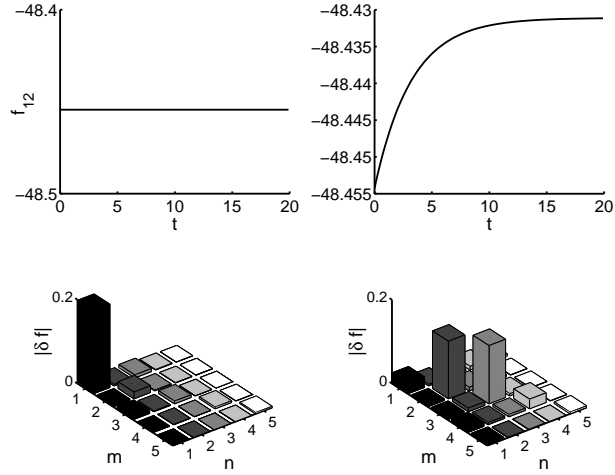}
\caption{The perturbation $\lambda \delta f^s_{\omega}$ (bottom)
  is shown in a rotated frame such that $f^{{\rm eq}}$ is diagonal;
  the resulting damped behavior (top) of the element $f_{12}$ of $f$
  in the box basis is due to the perturbation. The left and right
  figures correspond to the points marked (c) and (d) in
  Fig.~\ref{responseeigs}, respectively.}
\label{linresponse2}
\end{figure}

\begin{figure}[t]
\centering\includegraphics[scale=.35,angle=-90]{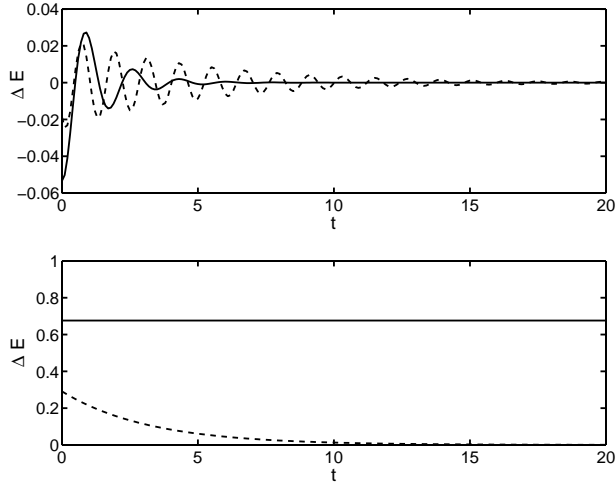}
\caption{The change in the total energy $\Delta
  E=E(f)-E(f^{{\rm eq}})$ as the system relaxes to its new
  equilibrium. In the top figure, the solid and dashed lines
  correspond to cases (a) and (b), respectively. Similarly in the
  bottom figure, the solid and dashed lines correspond to cases (c)
  and (d), respectively.}
\label{energycons}
\end{figure}

The real-time response is shown by plotting the off-diagonal matrix
element $f_{12}$ of the single particle density matrix in the box
basis as shown in Figs.~\ref{linresponse1} and \ref{linresponse2}.
In Fig. \ref{energycons}, we plot the change in the total energy
$\Delta E=E(f)-E(f^{{\rm eq}})$ as a function of time. In cases (a)
and (b) we see that the $\Delta E$ oscillates about zero and
eventually goes to zero. This is expected because such a
perturbation tends to create coherences, resulting in an energy change
by the amount of $E_{\rm{coh}}$ (coherence energy) that would
eventually decay down to zero. Similar damped behavior is
observed for case (d). Such an oscillatory damped behavior of the
total energy could be attributed to the Markov approximation in the
collision integral Eq.~(\ref{markovform}). On the other hand,
perturbations of the kind (c) increase the total particle number by
the amount $\delta N={\rm Tr}\{\lambda \delta f^{(s)}_{\omega}\}$ and
hence result in a finite change in total energy.

\section{Conclusion}
A non-Markovian version of the quantum kinetic theory is derived using
the prescription of a nonequilibrium statistical operator method as
outlined in Ref.~\cite{zubarev}. This theory is shown to conserve
energy in the $\eta \rightarrow 0$ limit.  Inclusion of quasi-particle
damping results in a description beyond the Born approximation with
energy conservation to $\phi^2$ order even in the Markov limit. To
obtain collision integrals that involve quasi-particle damping and
conserves energy precisely, one will have to calculate the $T$-matrix
in the full collision operator keeping terms of all orders in the
interaction.

We applied the generalized second order kinetic theory to the
nonhomogeneous dilute Bose gas confined in a spherical box to
numerically study the full non-equilibrium evolution of the system
towards equilibrium. The self-consistent distribution $f^{{\rm eq}}$
thus obtained is very close to the Bose-Einstein distribution as shown
in Fig.  \ref{linearfit}. We also observe a significant Hartree-Fock
self energy shift which depends on the single particle distribution
function $f$. The form of the damping function is important in
determining the line shape.  Particularly, the function with a
$1/\cosh$ type of behavior is found to be appropriate and gives
improved energy conservation due to smaller initial correlation
effect.

The importance of such a real-time calculation is apparent from the
full real-time response calculation, where we have calculated the
damping rates and frequencies. These damping rates correspond to a
shorter time scale compared to the equilibration time scale, which
depend on rates in and out of the various levels.

This simple model of a spherical trap can be easily extended to a more
realistic situation of a harmonic trap. As in
Refs.~\cite{walser,kane,martin}, the condensed component can be easily
included by introducing a symmetry broken mean-field,
$\alpha_i=\langle \hat{a}_i\rangle$, as one of the relevant
observables and Hartree-Fock-Bogoliubov quasi-particle excitations.
Such a calculation will allow us to make experimentally verifiable
predictions of damping rates of collective excitations. One can also
explore the possibility of including a time dependent potential or an
external force term to selectively excite one or more of the
collective modes. These possibilities will be explored in a
forthcoming article \cite{wachter2}.

\begin{acknowledgments} 
  We are extremely thankful to J. Cooper for the numerous and
  invaluable discussions. S.B. and M.H. acknowledge the financial
  support from the U. S. Department of Energy, Office of Basic Energy
  Sciences via the Chemical Sciences, Geosciences and Biosciences
  Division.  R.W. gratefully receives financial support from the
  Austrian Academy of Sciences through an APART grant.
\end{acknowledgments}
\appendix
\section{Reference distribution}
\label{appref}
The reference distribution $\sigma^{(0)}_{\{ f\}}$ of
Eq.~(\ref{boundaryc}) is parameterized through its expectation values
in Eq.~(\ref{averages}). From the structure of this quantum Gaussian
operator, it follows that
\begin{equation}
\sigma^{(0)}_{\{ {\cal K}(t,t_0)\,f\, {\cal K}^\dagger(t,t_0) \}}=   
{\hat U}^\dagger(t,t_0)\,
\sigma^{(0)}_{\{ f \}}\, {\hat U}(t,t_0).
\end{equation}
In the above, $\hat{U}$ represents the single particle propagator of
Eq.~(\ref{propagator}) acting in many-particle Fock-space and ${\cal K
  }$ is the corresponding single particle Hilbert-space propagator of
Eq.~(\ref{spprop}). This condition implies that
\begin{equation}
\partial_{\gamma_k}\sigma^{(0)} 
\text{Tr}\{[\hat{H}^{(0)},\hat{\gamma}_k]\}=
-[ \hat{H}^{(0)},\sigma^{(0)} ]
\end{equation}
and was used to obtain Eq.~(\ref{eqx}).

\section{Matrix elements}
\label{appmaxel}
The matrix element of Eq.~(\ref{maxel}) can be evaluated easily by
expanding the $\sin$ functions into co- and counter-propagating
complex exponents and by an additional partial integration. This
results in eight separate terms, i.e.,
\begin{eqnarray}
\phi^{ijkl}/ a_{\text{S}}&=&\frac{4}{\pi}
\int_0^\pi\sin{(i\, x)}\sin{(j\, x)}\sin{(k \,x)}\sin{(l\, x)}\,  
\frac{dx}{x^2},\nonumber\\
&=&F(i+j+k-l)+F(i+j-k+l)\nonumber\\
&+&F(i-j+k+l)+F(i-j-k-l)\nonumber\\
&-&F(i+j-k-l)-F(i-j+k-l)\nonumber\\
&-&F(i-j-k+l)-F(i+j+k+l),
\end{eqnarray}
where
\begin{eqnarray}
F(n)&=&\frac{1}{2\pi^2}[\cos{(n\pi)}+n\pi\, \text{Si}(n\pi)],\\
{\rm Si}(z)&=&\int_0^z\frac{\sin(t)}{t} \,dt.
\end{eqnarray}
An asymptotic expansion of the sine integral leads to the following
approximation that is correct at the $1\%$ level, i.e.,
\begin{eqnarray}
F(0)&=&\frac{1}{2\pi^2},\\
F(n>0)&\approx&\frac{1}{2\pi^2}
[\frac{\pi}{2} |n\pi|-\frac{\sin{(n \pi)}}{n \pi}
+\frac{2  \cos{(n\pi)}}{(n\pi)^2}].\nonumber
\end{eqnarray}


\begin{thebibliography}{}\label{sec:TeXbooks}

\bibitem{dalfovo}F. Dalfovo, S. Giorgini, L. Pitaevskii, and S. Stringari, Rev. Mod. Phys. {\bf 71}, 463 (1999).

\bibitem{murray}J. E. Williams and M. J. Holland, Nature {\bf 401}, 568 (1999).

\bibitem{haljan}B. P. Anderson, P. C. Haljan, C. E. Wieman, and E. A. Cornell, Phys. Rev. Lett. {\bf 85}, 2857 (2000).

\bibitem{raman}C. Raman, J. R. Abo-Shaeer, J. M. Vogels, K. Xu, and W. Ketterle, Phys. Rev. Lett. {\bf 87}, 210402 (2001). 

\bibitem{ketterle}J. R. Abo-Shaeer, C. Raman, and W. Ketterle, Phys. Rev. Lett. {\bf 88},  070409 (2002).

\bibitem{dalibard}K. W. Madison, F. Chevy, W. Wohlleben, and J. Dalibard, Phys. Rev. Lett. {\bf 84}, 806 (2000).

\bibitem{jin96}D. S. Jin, J. R. Ensher, M. R. Matthews, C. E. Wieman, and E. A. Cornell, Phys. Rev. Lett. {\bf 77}, 420 (1996).

\bibitem{edwards}M. Edwards {\sl et al.}, Phys. Rev. Lett. {\bf 77}, 1671 (1996).

\bibitem{stringari}S. Stringari, Phys. Rev. Lett. {\bf 77}, 2360 (1996).

\bibitem{walser}R. Walser, J. Williams, J. Cooper, and M. Holland, Phys. Rev. A {\bf 59}, 3878 (1999).

\bibitem{gardiner971}C. W. Gardiner and P. Zoller, Phys. Rev. A {\bf 55}, 2902 (1997).

\bibitem{gardiner973}C. W. Gardiner and P. Zoller, Phys. Rev. A {\bf 58}, 536 (1998).

\bibitem{milena}M. Imamovi\'c- Tomasovi\'c and A. Griffin, Phys. Rev. A {\bf 60}, 494 (1999).

\bibitem{gardiner975}C. W. Gardiner and P. Zoller, Phys. Rev. A {\bf 61}, 033601  (2000).

\bibitem{stoof}U. Al Khawaja and H. T. C. Stoof, Phys. Rev. A {\bf 62}, 053602  (2000).

\bibitem{morgan}S. A. Morgan J. Phys. B {\bf 33}, 3847 (2000).

\bibitem{burnett}T. K\"ohler and K. Burnett, Phys. Rev. A {\bf 65}, 033601  (2002).

\bibitem{morozov} V. G. Morozov and G. R\"opke, Ann. Phys. {\bf 278}, 127 (1999).

\bibitem{morozov2} V. G. Morozov and G. R\"opke, J. Stat. Phys. {\bf 102}, 285 (2001).

\bibitem{bonitz}M. Bonitz and D. Kremp, Phys. Lett. A {\bf 212}, 83 (1996).

\bibitem{kremp}D. Kremp {\sl et al.}, Physica B {\bf 228}, 72 (1996).

\bibitem{bonitzbook} M. Bonitz, {\sl Quantum Kinetic Theory} (B. G. Teubner Stuttgart, Leipzig, 1998).

\bibitem{zubarev} D. Zubarev, V. Morozov, and G. R\"opke, {\sl Statistical Mechanics of Nonequilibrium Processes} (Akademie Verlag, Berlin, 1996).

\bibitem{peletminskii}A. I. Akhiezer and S. V. Peletminskii, {\sl Methods of Statistical Physics} (Pergamon Press, Oxford, 1981).

\bibitem{chapman}S. Chapman and T. G. Cowling, {\sl The Mathematical Theory of Non-Uniform Gases} (Cambridge University Press, Cambridge, 1970).
 
\bibitem{haug}H. Haug and L. Banyai, Solid State Comm. {\bf 100}, 303 (1996).

\bibitem{semkat}D. Semkat and M. Bonitz in {\sl Progress in Nonequilibrium Green's Function}, edited by M. Bonitz (World Scientific, Singapore, 2000).

\bibitem{cornell}M. H. Anderson, J. R. Ensher, M. R. Matthews, C. E. Wieman, and E. A. Cornell, Science {\bf 269}, 198 (1995).

\bibitem{jin97}D. S. Jin, M. R. Matthews, J. R. Ensher, C. E. Wieman, and E. A. Cornell, Phys. Rev. Lett. {\bf 78}, 764 (1997).

\bibitem{walser2}R. Walser, J. Cooper, and M. J. Holland, Phys. Rev. A {\bf 63}, 013607 (2000).

\bibitem{kadanoff}L. Kadanoff and  G. Baym, {\sl Quantum Statistical Mechanics} (W. A. Benjamin, Inc., New York, 1962). 

\bibitem{kane}J. W. Kane and L. Kadanoff, J. Math. Phys. {\bf 6}, 1902 (1965). 
\bibitem{martin} P. C. Hohenberg and P. C. Martin, Ann. Phys. (N.Y.) {\bf 34}, 291 (1965).

\bibitem{wachter}J. Wachter, R. Walser, J. Cooper, and M. J. Holland, Phys. Rev. A {\bf 64}, 053612 (2001). 

\bibitem{wachter2}J. Wachter, S. Bhongale {\sl et al.}, in preparation.



\end{thebibliography}
\end{document}